# Terminal Performance of Lead-Free Pistol Bullets in Ballistic Gelatin Using Retarding Force Analysis from High Speed Video

Elijah Courtney, Amy Courtney, Lubov Andrusiv, and Michael Courtney
Michael_Courtney@alum.mit.edu

**Abstract**

Due to concerns about environmental and industrial hazards of lead, a number of military, law enforcement, and wildlife management agencies are giving careful consideration to lead-free ammunition. The goal of lead-free bullets is to gain the advantages of reduced lead use in the environment while maintaining equal or better terminal performance. Accepting reduced terminal performance would foolishly risk the lives of military and law enforcement personnel. This paper uses the established technique of studying bullet impacts in ballistic gelatin to characterize the terminal performance of eight commercial off-the-shelf lead-free handgun bullets for comparison with earlier analysis of jacketed lead bullets. Peak retarding force and energy deposit in calibrated ballistic gelatin are quantified using high speed video. The temporary stretch cavities and permanent wound cavities are also characterized. Two factors tend to reduce the terminal performance of these lead-free projectiles compared to similar jacketed lead designs. First, solid copper construction increase barrel friction, which reduces muzzle velocity and energy, and thus reduces the ability of the bullet to exert damaging forces in tissue simulant. Second, the lower density of copper requires a longer bullet for a given mass and caliber, which reduces remaining powder volume in the brass cartridge case, which also tends to reduce muzzle velocity and energy. The results of the present study are consistent with earlier analysis showing that expansion is necessary to maximize the potential for rapid incapacitation of enemy combatants. In spite of some new non-expanding nose designs that moderately increase forces between bullet and tissue, the largest retarding forces and highest incapacitation potential requires expanding bullets which maximize frontal area.

**Introduction**

Most pistol and rifle ammunition contain some lead in the bullet itself and also in the primer (in the form of lead styphnate). The development of current ammunition was influenced not only by cost and ease of manufacture, but also the good and reliable performance of lead-based components. Over the past two decades, interest has grown in removing lead from duty and hunting ammunition due to environmental and health concerns. For example, the US Army issued the new M855A1 load with lead-free bullets to troops in Afghanistan. An Air Force study showed that transitioning to training ammunition with lead-free bullets and primers can reduce instructor exposure to lead by 70% in indoor ranges and 41% in outdoor ranges (Cameron, 2006). However, the priority for selection of lead-free bullets must be that they perform equal to or better than the lead-based bullets they replace.

Three physical mechanisms shown to contribute to the wounding and incapacitation potential of penetrating projectiles are the permanent cavity, the temporary cavity, and a ballistic pressure wave radiating outward from the projectile (Courtney and Courtney, 2012). Since the force between bullet and tissue is the only direct interaction, all three mechanisms originate with the retarding force between bullet and tissue (Peters, 1990). The permanent cavity is often described as being caused by crushing of tissue by the bullet as it penetrates, but it is actually caused by an intense stress field in the immediate vicinity of the passing projectile. These stress waves decay rapidly with distance to levels below tissue damage thresholds a short distance from the path of the bullet through tissue. The temporary cavity is caused by the retarding force accelerating tissue forward and outward from the bullet path until elasticity causes the tissue to spring back into place. The temporary cavity can contribute to incapacitation and wounding when stretching beyond tissue's elastic limit damages tissue and enlarges the permanent

# Terminal Performance of Lead-Free Pistol Bullets in Ballistic Gelatin Using Retarding Force Analysis from High Speed Video

cavity. In addition, stretching can also cause nerve damage, and impact of the temporary stretch cavity with the spine can cause injury and hasten incapacitation. Third, a ballistic pressure wave propagates outward from the bullet path through tissue and can also contribute to wounding and incapacitation (Courtney and Courtney, 2007a; Suneson et al. 1990a, 1990b; Krasja, 2009; Selman et al., 2011).

Calibrated 10% ballistic gelatin has become widely accepted as a homogeneous tissue simulant that adequately reproduces average retarding forces and penetration depths of projectiles in tissue. Consistency has been reported between performance in calibrated 10% ballistic gelatin and autopsy findings (Wolberg, 1991). It is common to use ballistic gelatin to quantify penetration depth, permanent cavity, and temporary cavity volumes. It has recently been demonstrated that high speed video can be used to quantify the retarding force curve between bullet and tissue (Gaylord, 2013). Bo Janzon and colleagues seem to have originated the work of quantifying the retarding force and emphasizing its importance in wound ballistics (Janzon, 1983; Bellamy and Zajtchuk, 1990).

Since the target is deformed by the force being applied by the projectile, it is understood that work is being done on the target by the projectile. The energy lost by the projectile can be related to the force on the projectile by the Work-Energy Theorem. Using Newton's third law, the force of the bullet on the target, which is equal to the retarding force of the target on the bullet, can be obtained.

A high speed camera can record the location of the bullet as a function of time, which is required to quantify the bullet-tissue interaction using the Work-Energy Theorem. The path of the projectile is recorded from just prior to impact until the projectile comes to rest or exits the gelatin. The position of the projectile can then be tracked. With an appropriate scale in the plane of the gelatin surface, pixel identification can be converted to position, and then the change in position between subsequent frames of the video can be computed. This computation yields velocity of the projectile which decreases with time, from which acceleration can be calculated. The change in kinetic energy can also be computed. In the video, temporary and permanent cavities are also visible.

In this study, the wounding potential of eight commercial off-the-shelf lead-free handgun bullets was analyzed using the high speed video method to determine retarding force vs. penetration depth. The results were compared with earlier work (Gaylord et al., 2013) in four 9mm NATO loads. It is notable that the video method above also allows the energy deposited as a funtion of penetration depth to be computed, when velocity is expressed as a function of distance rather than time. This method can be used to estimate the volume/diameter of wounded tissue as a function of penetration depth (Janzon, 1983). It can also be used to estimate the probability of incapacitation given a hit, P(I/H) on an enemy combatant (Neades and Prather, 1991). The details of the mathematical model the Ballistics Research Laboratory (BRL) and later the Army Research Laboratory (ARL) have used to estimate conditional incapacitation probabilities, P(I/H), have changed slightly over the years but are consistently based on knowing the energy deposit as a function of penetration depth. The approach in this study employs the energy lost by the bullet in the first 15 cm of penetration, $E_{15}$, and the incapacitation probability model of Bruchey and Sturdivan (1968).

**Method**
The experimental and analysis methods as described in Gaylord et al. (2013) and Keys et al. (2015) were applied on a number of different lead-free bullet loads. Rounds in 9mm NATO, .357 SIG, and .40 S&W were fired from an appropriately chambered SIG P229 pistol. Rounds in .45 ACP were fired from a Kimber model 1911 (full size). Ammunition availability limited the number of loads tested to eight commercial off-the-shelf (COTS) products. The three 9mm NATO loads tested were an 85 grain DRT load (Dynamic Research Technologies), a 105 grain ME load (Lehigh Defense), and the Corbon loading of the 115 grain Barnes DPX. The two .357 SIG loads tested were the Lehigh Defense 115 grain Xtreme Penetrator Ammunition (XPA) and the Corbon loading of the 125 grain Barnes DPX. The .40 S&W load

# Terminal Performance of Lead-Free Pistol Bullets in Ballistic Gelatin Using Retarding Force Analysis from High Speed Video

tested was the Corbon loading of the 140 grain Barnes DPX. The two .45 ACP loads tested were the Corbon loading of the 160 grain Barnes DPX and the Magtech load using a 165 grain solid copper hollow point (SCHP).

The original published analysis method (Gaylord et al., 2013) applied Newton's laws and the Work-Energy Theorem based on the assumption that bullet mass was constant, which is appropriate for non-fragmenting bullets. This study applies the video analysis method for the first time to a fragmenting bullet (the 9mm NATO 85 grain DRT). The Appendix describes the appropriate modifications in the analysis method for fragmenting bullets. In cases of changing mass, a simple empirical model estimating how the bullet mass likely changes from its initial to final value provides a reasonable approximation of the retarding force.

Shortly after our testing was complete, an ammunition vendor published high speed videos of gelatin impacts for a number of additional lead-free bullets that were not available for our testing. Since those videos meet the technical requirements of adequate frame rate, resolution, horizontal scale, and motion in a plane perpendicular to the camera, the authors are planning to analyze those videos and publish a follow-up paper on terminal performance of an expanded list of COTS lead-free handgun bullets.

**Results**

The results are presented below in order of increasing caliber, beginning with the 9mm NATO loads and progressing through the .357 SIG, .40 S&W, and .45 ACP loads, respectively, with figures analogous to the previous two papers (Gaylord et al., 2013; Keys et al., 2015) showing the retarding force curves, temporary cavities, and permanent cavities for each load as determined by analysis of high speed video. After the retarding force curves, a summary table is presented showing estimates based on the BRL model for the probability of incapacitation given a hit. For ease of comparison, results from the previous two papers are included in the table for jacketed lead bullets in 9mm NATO and .357 SIG.

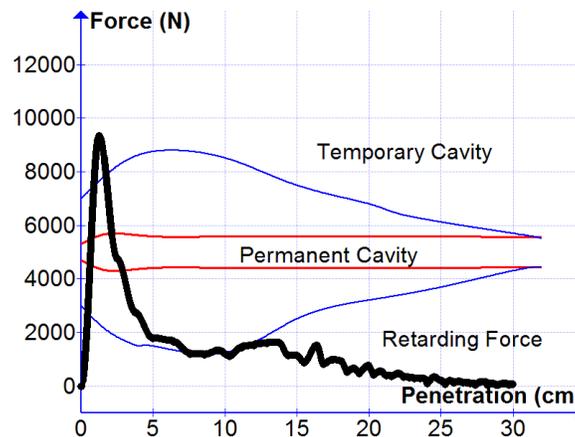

*Figure 1: Retarding force vs. penetration depth for the 9mm NATO 115 grain DPX impacting at 1157 ft/s. The permanent and temporary cavities are also shown. The vertical and horizontal length scales are the same.*

Figure 1 shows the force curve, temporary cavity, and permanent cavity of the 115 grain DPX in 9mm NATO. This is a solid copper bullet with a deep hollow point, which is scored so that the bullet expands symmetrically with six petals. The bullet remained nose forward when penetrating the gelatin (no tumbling), the peak retarding force was estimated to be near 9300 N, and the expanded bullet diameter was 0.70 inches. This peak retarding force is comparable to the largest 9mm NATO peak retarding force (9100 N) observed for the 127 grain Winchester Ranger SXT JHP design in the earlier study (Fig. 7 of Gaylord et al., 2013), even though both the impact energy and velocity are smaller. The high retarding

# Terminal Performance of Lead-Free Pistol Bullets in Ballistic Gelatin Using Retarding Force Analysis from High Speed Video

force is attributable to the rapid expansion of the scored hollow point in the solid copper DPX. The lighter bullet decelerates quickly and the large retarding force is not sustained through much penetration. Both the temporary cavity and permanent cavity of the 127 grain Winchester SXT are significantly larger than the 115 grain DPX in 9mm NATO.

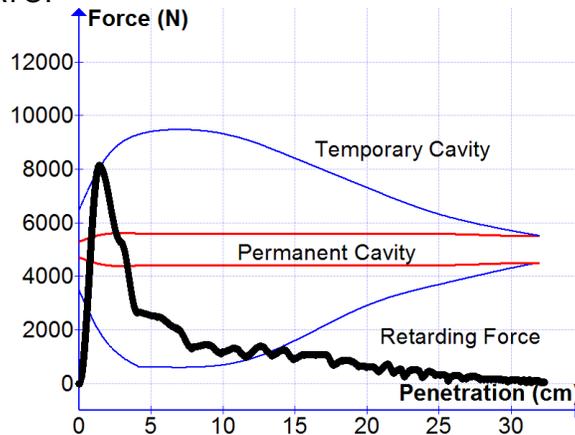

Figure 2: Retarding force vs. penetration depth for the 9mm NATO 105 grain ME impacting at 1225 ft/s. The permanent and temporary cavities are also shown. The vertical and horizontal length scales are the same.

The retarding force curve, temporary cavity, and permanent cavity of the 105 grain ME in 9mm NATO are shown in Figure 2. The peak retarding force is close to 8000 N, but the temporary cavity remains closer to its peak diameter (12.5 cm) more deeply into the penetration than for the 115 grain DPX. The expanded diameter is very close to that of the 115 grain DPX at 0.70", but since the solid copper 105 grain ME only has four petals, the frontal area is smaller when fully expanded. The 105 grain ME bullet remained nose forward until it exited the gelatin after 32 cm of penetration.

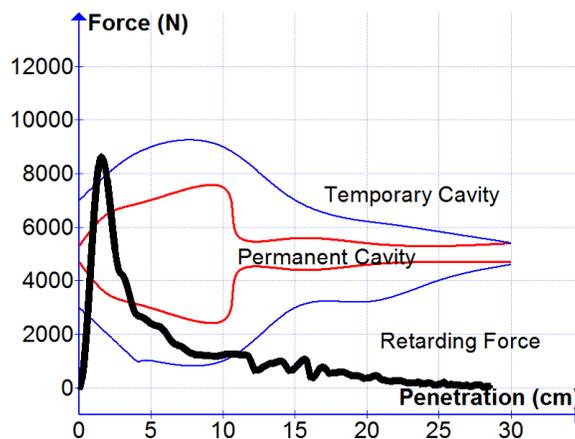

Figure 3: Retarding force vs. penetration depth for the 9mm NATO 85 grain DRT impacting at 1258 ft/s. The permanent and temporary cavities are also shown. The vertical and horizontal length scales are the same.

Figure 3 shows the force curve, temporary cavity, and permanent cavity of the 85 grain DRT in 9mm NATO. In contrast to the other lead-free bullets in 9mm NATO, this bullet is not solid copper; it is a jacketed hollow point with a copper alloy jacket and a soft core of other metals that expands and fragments as the bullet penetrates. Given the internal pressure specifications of the cartridge, velocities much higher than the measured 1258 ft/s should be attainable with an 85 grain bullet. The unusual shapes of the permanent and temporary cavities can be attributed to the fragmenting bullet and the tumbling of the

# Terminal Performance of Lead-Free Pistol Bullets in Ballistic Gelatin Using Retarding Force Analysis from High Speed Video

main fragment that penetrates most deeply. The first 11 cm of the permanent cavity is wide and shallow, due to the wide dispersion of bullet fragments before they come to rest, stopping before they penetrate 11 cm.  The peak retarding force near 8600 N is about 8% larger than for a non-fragmenting bullet that demonstrated the same deceleration in gelatin (see Appendix.)  Bullets with such broad and shallow wound profiles are generally not recommended for law enforcement duty use, because vital organs are usually located deeper from commonly encountered shot angles.  This bullet also fails to meet the FBI recommended minimum penetration depth of 12 inches for duty use.

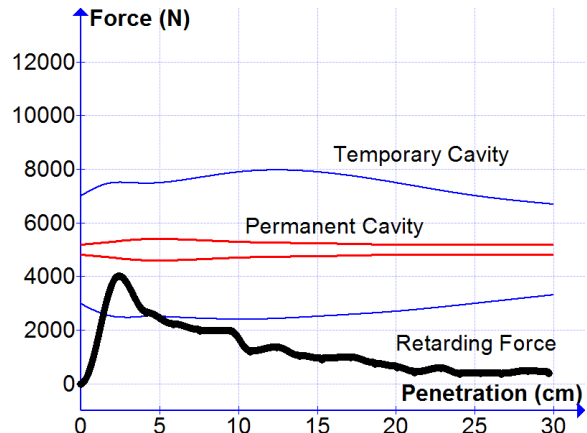

*Figure 4: Retarding force vs. penetration depth for the 115 grain Lehigh Defense XPA in .357 SIG impacting at 1396 ft/s. The permanent and temporary cavities are also shown.  The vertical and horizontal length scales are the same.*

The force curve, temporary cavity, and permanent cavity of the 115 grain XPA in .357 SIG are shown in Figure 4.  This bullet design attempts to improve performance over other non-expanding bullet designs with machined flutes in the nose of the bullet to redirect tissue away from the bullet path and increase the retarding force.  However, comparing these results with the results for a 125 grain flat nosed FMJ (Figure 1, Keys et al., 2015) shows they both have a peak retarding force close to 4000 N.  However, the 115 grain XPA does have a temporary cavity with a larger peak diameter (8.3 cm) than the flat nosed FMJ (7 cm).  This bullet exited the 33 cm long (13 inches) gelatin block with over 900 ft/s of velocity, so it would likely penetrate much deeper.  Also, because of the solid copper design with no hollow point and scoring in the nose, the XPA would likely excel at penetrating intermediate barriers en route to the target.  Figure 4 illustrates the inherent limitations of non-expanding bullets for producing the large retarding forces necessary to hasten incapacitation with hits to soft tissue.

Figure 5 shows the force curve, temporary cavity, and permanent cavity of the 125 grain DPX in .357 SIG.  The peak retarding force of 11,100 N compares favorably with all the lead-free bullets in the present study (only the 165 grain MagTech in .45 ACP is larger), as well as lead-based 9mm and .357 SIG bullets in prior studies (Gaylord et al., 2013; Keys et al., 2015).  The peak diameter of the temporary cavity is 14 cm, and the bullet remained nose forward throughout penetration until exiting the 33 cm long block with just over 100 ft/s residual velocity.  The expanded diameter of 0.71 inch is slightly smaller than that of the 125 grain (lead-based) Winchester Ranger SXT load which measured 0.775 inch in diameter (Keys et al., 2015).  This load may provide the highest level of lead-free terminal performance available in pistols chambered for 9mm NATO, .357 Sig or .40 S&W.

# Terminal Performance of Lead-Free Pistol Bullets in Ballistic Gelatin Using Retarding Force Analysis from High Speed Video

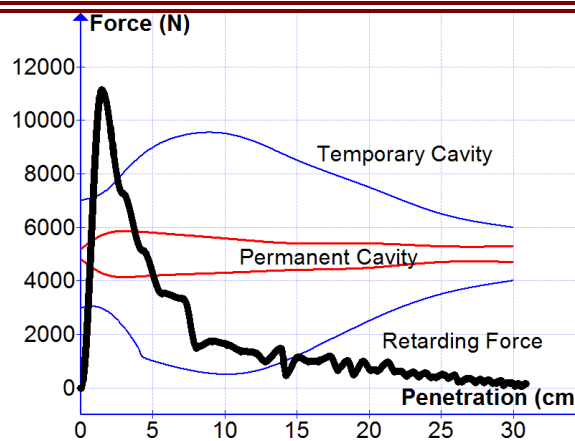

Figure 5: Retarding force vs. penetration depth for the 125 grain Barnes DPX load in .357 SIG impacting at 1272 ft/s. The permanent and temporary cavities are also shown. The vertical and horizontal length scales are the same.

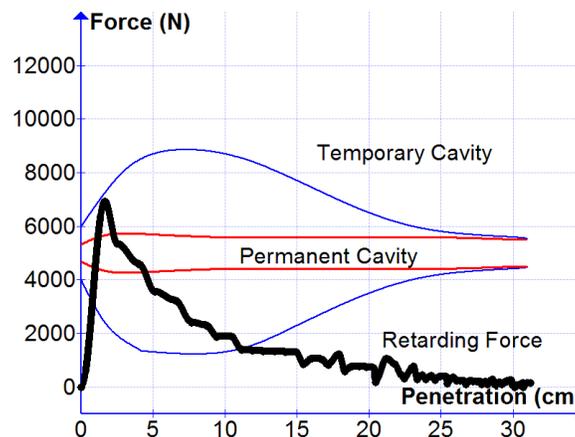

Figure 6: Retarding force vs. penetration depth for the 140 grain Barnes DPX load in .40 S&W impacting at 1084 ft/s. The permanent and temporary cavities are also shown. The vertical and horizontal length scales are the same.

The force curve, temporary cavity, and permanent cavity of the 140 grain DPX in .40 S&W are shown in Figure 6. At under 7000 N, the peak retarding force is smaller than most peak forces of expanding hollow points in 9mm NATO and .357 SIG. All three Corbon loads of the different Barnes DPX bullets underperformed relative to their muzzle velocity and energy specifications (as claimed by Corbon). Specifically, this bullet impacted with only 365 ft lbs of energy (instead of the advertised 448 ft lbs), which hindered its performance in this study. It is notable that, unlike all the lead-free 9mm NATO and .357 SIG bullets, this bullet tumbled after expanding and before coming to rest after 31 cm of penetration.

Figure 7 shows the force curve, temporary cavity, and permanent cavity of the 160 grain DPX in .45 ACP. The unusual shapes of the temporary and permanent cavities are caused by the bullet tumbling after it expands. The peak retarding force is near 9200 N, but unlike less energetic bullets where the retarding force has markedly declined by the time the bullet has penetrated 5 cm, the mass, frontal area, and retained velocity provide a retarding force above 5000 N at that point. 45 caliber bullets have the advantages of unexpanded diameter, expanded diameter, and mass, but in many cases their terminal performance is hindered by low velocities. This bullet impacts at 1185 ft/s and expands to 0.75 inches in diameter.

# Terminal Performance of Lead-Free Pistol Bullets in Ballistic Gelatin Using Retarding Force Analysis from High Speed Video

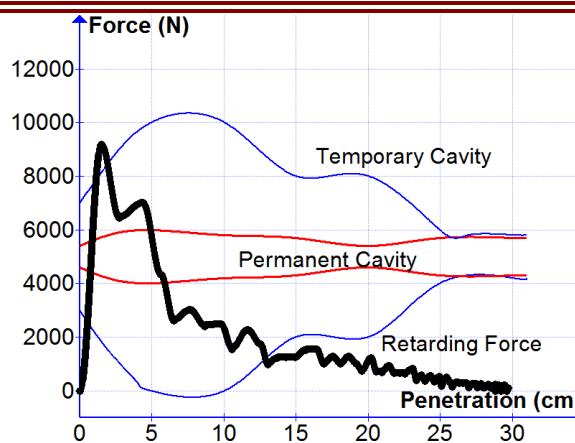

*Figure 7: Retarding force vs. penetration depth for the 160 grain Barnes DPX load in .45 ACP impacting at 1185 ft/s. The permanent and temporary cavities are also shown. The vertical and horizontal length scales are the same.*

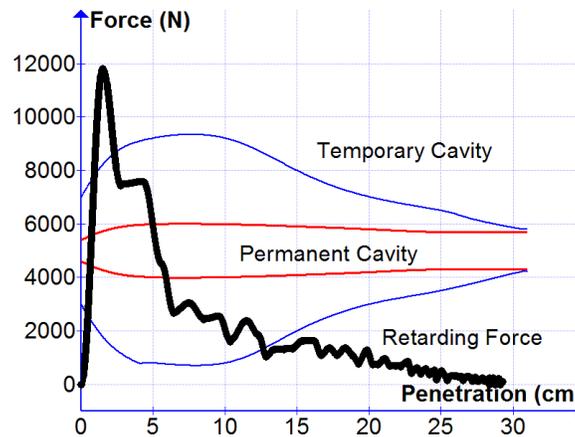

*Figure 8: Retarding force vs. penetration depth for the 165 grain Magtech solid copper hollow point load in .45 ACP impacting at 1222 ft/s. The permanent and temporary cavities are also shown. The vertical and horizontal length scales are the same.*

As described in the introduction, this video analysis method can be applied to determine the energy deposited to the target in any desired range of penetration depths. Energy deposit can be used, for example, to estimate the volume/diameter of wounded tissue as a function of penetration depth (Janzon, 1983). Over the years, the BRL and ARL have used several similar methods based on energy deposit as a function of penetration depth to estimate the probability of incapacitating an enemy combatant given a hit. Neades and Prather (1991) published a method to estimate the probability of incapacitation given a hit, P(I/H), on an enemy combatant based on the energy deposit in the first 15 cm of penetration ($E_{15}$). Implementation details for handgun bullets in gelatin are described in Gaylord et al. (2013).

Table 1 shows the resulting conditional incapacitation probabilities for the eight lead-free loads from the present study compared with the four lead-based 9mm NATO loads from Gaylord et al. (2013) and the four lead-based .357 SIG loads from Keys et al. (2015). The best-performing lead-based load in 9mm NATO performed better than all three lead-free loads in 9mm NATO, with a P(I/H) of 0.515 compared with P(I/H) values ranging from 0.330 to 0.460 for the lead-free bullets. Likewise, the best-performing lead-based load in .357 SIG performed better with a P(I/H) of 0.601 compared with the lead-free loads in .357 SIG which had P(I/H) values ranging from 0.390 to 0.546. The lead-free load in .40 S&W also underperformed compared to the lead-based loads in .357 SIG and 9mm NATO.

# Terminal Performance of Lead-Free Pistol Bullets in Ballistic Gelatin Using Retarding Force Analysis from High Speed Video

*Table 1: Conditional incapacitation probabilities P(I/H) for lead-free bullets using the BRL method (Bruchey and Sturdivan, 1968; Neades and Prather, 1991). The range of equivalent effectiveness for the M16 (using the M193 bullet) is also shown (M16, 1968). The lead-based bullet results are from Gaylord et al. (2013) and Keys et al.(2015)*

| Load | $E_{15}$ | P(I/H) | M193 Equivalent |
|---|---|---|---|
|  | (ft lbs) |  | Range (yards) |
| **9mm NATO** | Lead-Free |  |  |
| 115 DPX | 289 | 0.460 | 460 |
| 105 ME | 290 | 0.461 | 460 |
| 85 DRT | 269 | 0.441 | 470 |
| **.357SIG** | Lead-Free |  |  |
| 115 XPA | 222 | 0.390 | 530 |
| 125 DPX | 389 | 0.546 | 340 |
| **.40 S&W** | Lead-Free |  |  |
| 140 DPX | 305 | 0.475 | 420 |
| **.45 ACP** | Lead-Free |  |  |
| 160 DPX | 431 | 0.577 | 290 |
| 165 SCHP | 473 | 0.605 | 270 |
| **9mm NATO** | Lead-Based |  |  |
| 124 FMJ | 172 | 0.330 | 620 |
| 147 WWB | 197 | 0.361 | 570 |
| 147 SXT | 240 | 0.410 | 490 |
| 127 SXT | 350 | 0.515 | 380 |
| **.357 SIG** | Lead-Based |  |  |
| 125 FMJ | 222 | 0.390 | 530 |
| 125 SXT | 466 | 0.600 | 270 |
| 115 GDHP | 467 | 0.601 | 270 |
| 125 FedHP | 437 | 0.581 | 290 |

Among the lead-free loads tested, only the two expanding solid copper bullets in .45 ACP performed comparably to the best lead-based bullets in 9mm NATO and .357 SIG, with P(I/H) values of 0.577 and 0.601 for the 160 grain DPX and the 165 grain SCHP, respectively.

**Discussion**
The highest P(I/H) value of any non-expanding bullet in Table 1 is 0.390, which approximates the effectiveness of the M193 bullet from an M16 rifle at 530 yards. Such a lack of terminal effectiveness is unacceptable, since the M16 family of rifles are rarely effective beyond 300 yards (Ehrhart, 2009), and most duty use of handguns is in situations where lives of US personnel are in imminent danger if the enemy combatant is not quickly neutralized. Poor performance of non-expanding bullets is consistent with the poor performance of all non-expanding handgun bullets in the Marshall and Sanow (2001)

# Terminal Performance of Lead-Free Pistol Bullets in Ballistic Gelatin Using Retarding Force Analysis from High Speed Video

epidemiological type analysis of shootings in humans as well as the long incapacitation times resulting from non-expanding handgun bullets in a laboratory study in live goats (Courtney and Courtney, 2007c).

In contrast, Table 1 shows three lead-free expanding bullets and four lead-based expanding bullets with P(I/H) values above 0.500 that are available in COTS loaded ammunition. The differences between expanding and non-expanding handgun bullet performance is much larger than the difference between lead-based and lead-free bullet designs.

However, it is also apparent that lead-free bullet designs tend to underperform lead-based designs, especially in the smaller calibers due to several design challenges for lead-free bullets. The design challenge of reliable expansion has been largely overcome through the use of a deep hollowpoint nose cavity and scoring or cutting the front section of the bullet to reliably expand out in four to six defined petals. However, for a given bearing surface area, solid copper bullets have much more barrel friction than jacketed lead bullets, and energy is expended overcoming barrel friction rather than increasing muzzle velocity. For example, Table A1 of Summer and Courtney (2014) shows that solid copper bullets in 5.56mm NATO tend to lose 350 - 750 ft lbs of energy to barrel friction; whereas, jacketed lead bullets of comparable length only lose 250 - 350 ft lbs of energy to barrel friction. Increased friction is further exacerbated by the lower density of copper which tends to increase the length and bearing surface of copper bullets by 10-20% compared with jacketed lead bullets of the same weight. Because of constraints on cartridge overall length, longer copper bullets also reduce the powder volume available in cartridge cases, which reduces the velocities that can be achieved for the same peak pressure constraints.

The solid copper bullets in .45 ACP have ameliorated these design challenges by using lighter and shorter solid copper bullets compared with the 200-230 grain jacketed lead bullets that are more typical of this cartridge. Solid copper bullets in the 160-165 grain range are short enough not to increase friction or intrude on the powder volume, so they allow high levels of velocity and kinetic energy. Yet the diameter of the cartridge and the strength of solid copper allow for a good combination of expanded diameter, retarding force, and penetration depth even at lower sectional densities of light-for-caliber bullets. The 45 caliber design also has the advantage that while surface area increases as diameter squared (increasing both area impacting the target and also the bottom area on which internal pressure pushes during acceleration), the bearing surface only increases as a linear function of the diameter. Thus, benefits of increased diameter accrue with the square of the diameter, but the disadvantage of increased friction only increases linearly.

In summary, this paper presents results of analysis of high speed video to quantify retarding forces and wound cavities of eight lead-free handgun bullets in duty calibers and compares them with results of eight lead-based bullets from earlier studies. In the smaller calibers (9mm NATO, .357 SIG, and .40 S&W), lead-free designs still have room for improvement to perform as well as the best available lead-based designs.

# Terminal Performance of Lead-Free Pistol Bullets in Ballistic Gelatin Using Retarding Force Analysis from High Speed Video

**Appendix: Mass Decay Model for Fragmenting Bullets**

Recalling Newton's Second Law in a form

$$F = \frac{d}{dt}(mv),$$

where $F$ is the retarding force, $m$ is the bullet mass, and $v$ is the bullet velocity which can be determined by analysis of the high speed video as described in Gaylord et al. (2013) and applied in Keys et al. (2015). Newton's second law as written $F = ma$ applies well in cases in which bullet mass remains constant (no fragmenting), since $dm/dt = 0$.

Applying the chain rule in cases of changing mass gives

$$F = \frac{d(mv)}{dt} = v\frac{dm}{dt} + m\frac{dv}{dt}.$$

Unfortunately, determining the mass as a function of time, $m(t)$, with a high level of confidence and accuracy while the bullet is penetrating the ballistic gelatin is prohibitively difficult. However, since we know the initial and final masses (by measuring the mass of the final core or the fragment that penetrates furthest in the ballistic gelatin), an empirical model can be constructed with the proper limiting behavior for the initial mass, $m_i$, and the final mass, $m_f$.

Here, we use a function of the form

$$m(t) = m_f + (m_i - m_f)f(t),$$

where $f(t)$ is an appropriately chosen decay function which goes from 1 to 0 as the bullet penetrates ensuring the proper initial and final masses. One might consider different functional forms for $f(t)$. For example, many decaying phenomena are accurately represented by a decaying exponential. However, that was deemed to be inappropriate here, since a decaying exponential suggests the largest mass loss rate ($dm/dt$) occurs at the moment of impact and declines monotonically through the penetration. Observation of many fragmenting bullets in gelatin shows that fragmenting bullets penetrate several cm before fragmentation begins, then have a region of rapid mass loss, then the core or heaviest fragment continues to penetrate more deeply after the smaller fragments have come to rest. This suggests a decaying function of the form

$$f(x) = \frac{1}{1+(x/d)^n},$$

where $x$ is the penetration depth, $d$ is the penetration depth at which half the mass has been lost, and $n$ is an appropriate exponent to correctly model the onset and peak slope of the mass loss. Note that it is convenient to convert the input of the decay function, $f(x)$, to penetration depth, $x$, rather than time. This is possible without loss of applicability or generality, because the chain rule can simply be used to convert between derivatives with respect to time and derivatives with respect to position. Expressing the decay in terms of penetration depth is convenient, because the final frames of the video (or still pictures of the gelatin) can be used to estimate $d$ and verify by inspection that the decay function represents the apparent mass loss in the gelatin.

Several possible choices of mass decay models are shown in Figure A1. Choosing $n = 1$ produces a function whose largest derivative is at $x = 0$, which is clearly unphysical, since bullets tend to penetrate several cm before fragmentation begins. The value of $n$ needs to be chosen to better approximate the onset of fragmentation. The other curves in the graph represent cases of $n = 2$ and *4*, and $d = 6$ cm and *10* cm. The value of $d$ is chosen to be a midpoint or effective average of where the fragments come to rest in the gelatin, and it can be inferred from inspection of the gelatin after the shot. Some controlled

# Terminal Performance of Lead-Free Pistol Bullets in Ballistic Gelatin Using Retarding Force Analysis from High Speed Video

fracturing or pre-fragmented rounds have more deeply penetrating fragments, so a value of *d* of *10* cm or greater might be appropriate. However, the fragments from the 85 grain DRT bullet from the present study (Figure 3) penetrated approximately 6 cm on average.

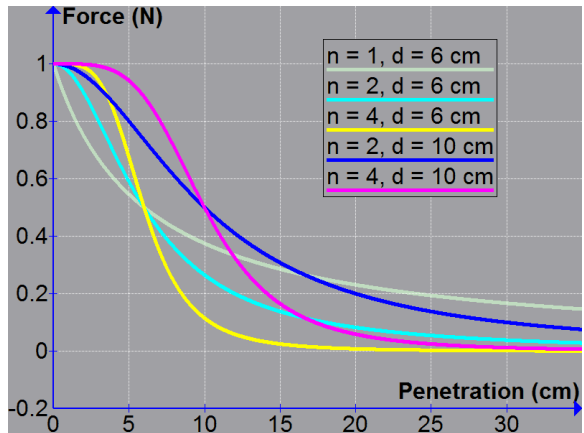

*Figure A1: Mass decay curves with different parameters for approximating fragmentation effects in retarding force models.*

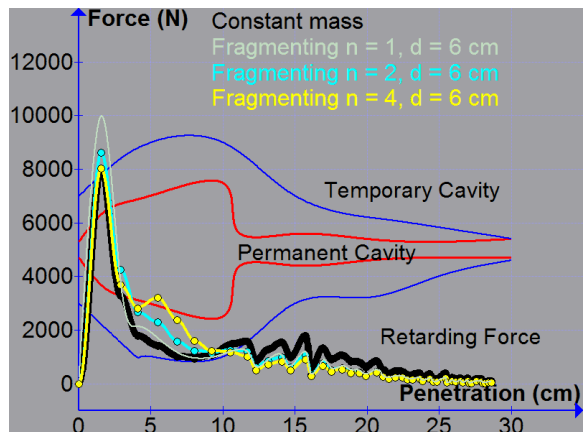

*Figure A2: Effects of different mass decay models on resulting retarding force curves.*

The possible effects of different exponents are shown in Figure A2 for the 85 grain DRT bullet in 9mm NATO, along with the results of neglecting fragmentation all together. This model used an initial mass of 85 grains and a final mass of 40.5 grains. Note that the effects of different mass decay parameters is only to move area under the force curve from one depth to another. As is well known, total area under the force curve is equal to the impact energy. If the mass decay model has mass being lost quickly and early in the penetration, the initial peak of the force curve will be increased, but the later part of the force curve will be reduced to compensate so that the total area under the curve remains unchanged. A mass decay model which has its peak in *dm/dt* (the mass loss rate) later will increase the force near the peak in *dm/dt* while maintaining constant area under the retarding force curve.

An exponent of *n = 1* produces the largest peak retarding force, because it maximizes (*dm/dt*) right at impact, which is very close to the same instant the acceleration (*dv/dt*) is the largest. The peak in the retarding force curve assuming constant mass (wrongly) is 7990 N. The peak in the retarding force curves assuming *n* = 1, 2, and 4 are 9970 N, 8610 N, and 8038 N, respectively. The difference between the resulting peaks for the two most physically reasonable estimates (*n* = 2 and *n* = 4) is only 7%. Admittedly,

# Terminal Performance of Lead-Free Pistol Bullets in Ballistic Gelatin Using Retarding Force Analysis from High Speed Video

we have little more than intuition and experience behind our choice of $n$ = 2 to produce the resulting Figure 3. A skeptical reader may object to the lack of rigor and certainty in our mass decay models, but Figure A2 shows that the main features and peak of the resulting retarding force curve are not very sensitive to the details of the mass decay model. The same mass decay model can also be used to compute a resulting retarding force curve by numerically estimating the retarding force as *dE/dx* along the penetration path. Our practice is to estimate retarding force curves by both methods (Newton's second law and *dE/dx*) and choose the curve whose area is closer to the change in kinetic energy along the path. The difference in peak force between the methods is typically around 5%.

**Acknowledgements**
This research was supported in part by BTG Research (www.btgresearch.org) and by the United States Air Force Academy. The views expressed in this paper are those of the authors and do not necessarily represent those of the U.S. Air Force Academy, the U.S. Air Force, the Department of Defense, or the U.S. Government. The authors are grateful to Louisiana Shooters Unlimited for providing range facilities where the experiments were performed as well as some of the ammunition and equipment.

# Terminal Performance of Lead-Free Pistol Bullets in Ballistic Gelatin Using Retarding Force Analysis from High Speed Video

# Terminal Performance of Lead-Free Pistol Bullets in Ballistic Gelatin Using Retarding Force Analysis from High Speed Video

Lubov Andrusiv
United States Air Force Academy
2354 Fairchild Drive
USAF Academy, CO 80840
lubov.andrusiv@usafa.edu

Elijah Courtney
BTG Research
9574 Simon Lebleu Road
Lake Charles, LA, 70607
elijahdscourtney@gmail.com

Amy Courtney
Exponent, Inc.
3440 Market Street
Philadelphia, PA 19104, USA
Amy_Courtney@post.harvard.edu

Michael Courtney
BTG Research
9574 Simon Lebleu Road
Lake Charles, LA, 70607
Michael_Courtney@alum.mit.edu